\documentclass[10pt,twocolumn,letterpaper]{article}

\usepackage{tabularx} 
\usepackage[numbers]{natbib}

\usepackage[spanish,english]{babel}
\usepackage[utf8x]{inputenc}
\usepackage[T1]{fontenc}
\usepackage{float}

\usepackage[a4paper,top=3cm,bottom=2cm,left=3cm,right=3cm,marginparwidth=1.75cm]{geometry}

\usepackage{amsmath}
\usepackage{graphicx}
\usepackage[colorinlistoftodos]{todonotes}
\usepackage[colorlinks=true, allcolors=blue]{hyperref}

\title{Establishment and Solution of a Multi-Stage Decision Model Based on Hypothesis Testing and Dynamic Programming Algorithm}

\usepackage{authblk}

\author[1]{Ziyang Liu}
\author[2]{Yurui Hu}
\author[3]{Yihan Deng}

	\affil[1]{\small{School of Future Science and Engineering,
Soochow University,
Suzhou, China,
ziyannn@yeah.net}}
\affil[2]{School of Future Science and Engineering,
Soochow University,
Suzhou, China,
1462362144@qq.com}
\affil[3]{School of Future Science and Engineering,
Soochow University,
Suzhou, China,
76581619@qq.com} 

\date{} 

\begin{document}

\maketitle

\selectlanguage{english}
\begin{abstract}
This paper introduces a novel multi-stage decision-making model that integrates hypothesis testing and dynamic programming algorithms to address complex decision-making scenarios.Initially,we develop a sampling inspection scheme that controls for both Type I and Type II errors using a simple random sampling method without replacement,ensuring the randomness and representativeness of the sample while minimizing selection bias.Through the application of hypothesis testing theory,a hypothesis testing model concerning the defect rate is established,and formulas for the approximate distribution of the sample defect rate and the minimum sample size required under two different scenarios are derived. Subsequently,a multi-stage dynamic programming decision model is constructed.This involves defining the state transition functions and stage-specific objective functions,followed by obtaining six optimal decision strategies under various conditions through backward recursion.The results demonstrate the model’s potent capability for multi-stage decision-making and its high interpretability,offering significant advantages in practical applications.
\end{abstract} 

\section*{I.Introduction}
Decision-making in complex environments often requires a methodical approach to handle uncertainties and dynamically changing conditions. Traditional decision models frequently fall short when faced with multi-stage scenarios where decisions at one stage influence outcomes in subsequent stages. To address this gap, we propose a comprehensive model that combines hypothesis testing with dynamic programming to formulate and solve multi-stage decision problems. This approach allows for a more nuanced control of decision-making errors and adapts to the evolving nature of decision scenarios.

Zhu Chenlu and colleagues \cite{Zhu2024} initially employed probabilistic interval hesitant fuzzy sets for depicting the decision-making system’s hesitancy, followed by the development of an optimization model leveraging the score function, deviation function, and information entropy of these fuzzy elements to extract probabilistic information. Subsequently, to enhance the computational efficiency, a cloud model was adopted, facilitating the transformation from PIVHFS to this cloud model through an established optimization framework. Liu Weiqiao and Zhu Jianjun \cite{Liu2022} introduced a multi-stage decision-making approach that incorporates quantum guidance, utilizing the normal cloud model for expert relocation. Zhong Xiangyu and associates \cite{Zhong2021} developed a novel consensus model for large-group multi-attribute decision-making, asserting the equal significance of basic and ordinal consensus by integrating these aspects. Mao Xingyun and team \cite{Mao2025} introduced a preprocessing strategy aimed at simplifying the computational demands of the latent group lasso (LGL) issue, which notably reduces unnecessary overlapping support groups while maintaining a constant count of essential support groups. Li Lubo and collaborators \cite{Li2024} tackled the Manufacturing Project Scheduling Problem with Preparation Time in Dynamic and Disruptive Environments (MPSPST-DIE), a scenario frequently arising in uncertain and spatially disturbed production settings, by crafting a multi-agent genetic planning super heuristic (HH-MGP) algorithm for its resolution.

Hypothesis testing, a statistical method used to make inferences about populations based on sample data, provides a structured framework to control and assess the risk of incorrect decisions (Type I and Type II errors). By applying these principles, we ensure that the decision-making process is both evidence-based and statistically sound. The incorporation of dynamic programming, a mathematical optimization method, allows for the efficient solving of complex problems by breaking them down into simpler, interrelated stages. This method’s forward-looking nature and its ability to consider the entirety of a decision path make it ideally suited for multi-stage decision-making.

The synergy between hypothesis testing and dynamic programming in our model facilitates the handling of decision-making scenarios with a higher degree of complexity and uncertainty. By systematically addressing the statistical and sequential aspects of decision-making, our model is capable of providing robust solutions that are both optimal and practical.

\section*{II.Design of sampling test methods}

\textbf{A.Hypothesis testing and approximate distribution}

Hypothesis testing

Hypothesis testing is the process of statistically analyzing sample data to determine whether a hypothesis (usually about an overall parameter) is valid. In this problem, the supplier claims that the defective rate of spare parts will not exceed a certain nominal value, and hypothesis testing can be used to assess whether the sample data supports this claim by the supplier.

In this problem, the following two hypotheses can be constructed:

Original assumption $H_0:p\le p_0$, the rate of defective products does not exceed the nominal value claimed by the supplier;

Alternative hypothesis $H_1:p\ge  p_0$, the rate of defective products exceeds the nominal value claimed by the supplier.

Approximate distribution

Let the total number of spare parts be , in which each spare part can only be qualified or non-qualified products, so we can specify the following overall unit index:

\begin{equation}
           Y_{i} = 
\begin{cases} 
1,& \text{If the spare parts are not qualified,} \\
0,& \text{If the spare parts are qualified.}
\end{cases}
       \end{equation}

From the above only 0, 1 two kinds of index value of the overall take a sample size of n simple random sample, let  be the number of substandard spare parts in the sample, then the sample defective rate:

\begin{equation}
    p=\frac{a}{n}
\end{equation}

Which is the sample mean, and thus the sample variance:

\begin{equation}
   s^{2}=\frac{1}{n-1}\cdot \sum_{i=1}^{n}\cdot \left ( y_{i}-\bar{y}   \right )^{2} =\frac{n}{n-1}\cdot p\cdot \left ( 1-p \right )
\end{equation}

And the variance of the sample defective rate is:

\begin{equation}
    v\left(p\right)=\frac{1-f}{n}\cdot s^{2}=\frac{1-f}{n-1}\cdot p\cdot \left( 1-p \right) 
\end{equation}

Since no-return simple random sampling is used $a=n\cdot p$,  obeys the hypergeometric distribution:

\begin{equation}
    a \sim \text{HG}(N,M,n  )
\end{equation}

The expectation of the hypergeometric distribution is:

\begin{equation}
    E(a)=n\cdot \frac{M}{N}
\end{equation}

And the variance of the hypergeometric distribution is:

\begin{equation}
    V(a)=n\cdot \frac{M}{N}\cdot \left ( 1-\frac{M}{N}  \right ) \cdot \frac{N-n}{N-1}
\end{equation}

where $M$ is the number of nonconforming parts in the population $N$. It is useful to note that $P=\frac{M}{N}$, and $P$ is the actual defective rate. $\frac{N-n}{N-1}$ is the correction factor due to non-return sampling, as distinguished from the binomial distribution.

When $N$ is large, $\frac{N-n}{N-1}\approx 1$ , and $a$ approximately follows the binomial distribution:

\begin{equation}
    a \sim \text{B}(n, P)
\end{equation}

When $n$ is also large, $a$ again approximately follows a normal distribution:

\begin{equation}
    a \sim N(n P , n P \left ( 1-P  \right ) )
\end{equation}

In turn, the distribution of the sample defective rate $p$ can be obtained from equation (2):

\begin{equation}
    p \sim N\left(P, \frac{P(1 - P)}{n}\right)
\end{equation}

\noindent\textbf{B.Sampling and testing programs to control Type I errors}

In this paper, we wish to design a sampling scheme that can determine whether to reject the lot of spare parts at a 95\% confidence level. In this case, the firm wants to control that the probability of the first type of error (wrongly rejecting the lot of spare parts when $H_0$ is true) $\alpha$ does not exceed 0.05.

Test statistic

The sample defect rate $p$ obeys a normal distribution if the actual defect rate $P=p_0$:

\setcounter{equation}{0}

\begin{equation}
    p \sim N\left(p_0, \frac{p_0(1 - p_0)}{n}\right)
\end{equation}

In order to check whether the defective rate provided by the supplier exceeds the nominal value $p0$, we standardize the sample defective rate $p$ to obtain the test statistic $Z$:

\begin{equation}
    Z = \frac{\hat{p} - p_0}{\sqrt{\frac{p_0(1 - p_0)}{n}}}
\end{equation}

At this point, $Z$ obeys the standard normal distribution $N(0, 1)$. From the standard normal distribution table, the critical value for the 95\% confidence level is:

\begin{equation}
    Z_{0.95} = 1.645
\end{equation}

To derive the minimum sample size, we set the permissible absolute error limit to $d$, i.e., we want the difference between the defective rate in the sample and the nominal defective rate $p_0$ to be no more than $d$. At this point, the

\begin{equation}
    Z_{0.95} = \frac{d}{\sqrt{\frac{p_0(1 - p_0)}{n}}}
\end{equation}

The final formula to obtain the sample size is:

\begin{equation}
    n = \frac{p_0(1 - p_0)}{\left( \frac{d}{Z_{0.95}} \right)^2}
\end{equation}

 $d$ is in the range of 0.02 to 0.09, the sample size at each step of 0.01, varies as shown in Figure \ref{fig:Change in sample size n for different values of d}:

 \begin{figure}[H]
  \centering
  \includegraphics[width=0.4\textwidth]{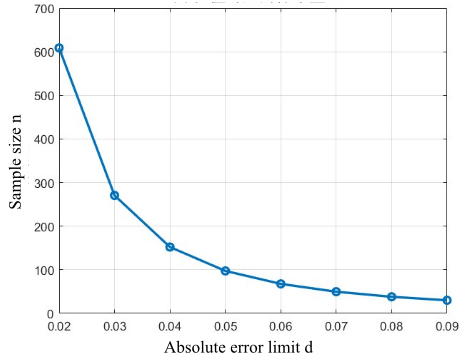}
  \caption{Change in sample size n for different values of d}
  \label{fig:Change in sample size n for different values of d}
\end{figure}

\noindent\textbf{C.Sampling and testing programs to control Type II errors}

Enterprises wish to decide whether or not to accept the lot of spare parts at the 90\% efficacy level. The second type of error risk is the probability of incorrectly accepting a lot that exceeds the defect rate when $H1$ is true.

For this reason, this paper sets up an efficacy model with efficacy of $1 - \beta = 90\%$, i.e., the probability of failing to reject the null hypothesis (Type II error probability) is expected to be no more than $\beta = 0.1$.

Assuming that the actual defective rate is $p_1$, the defective rate $p$ in the sample still follows an approximately normal distribution:

\setcounter{equation}{0}

\begin{equation}
    p \sim N\left(p_1, \frac{p_1(1 - p_1)}{n}\right)
\end{equation}

We wish to detect the difference between the actual defective rate $p_1$ and the nominal defective rate $p_0$. In the efficacy analysis, the test statistic $Z$ is constructed:

\begin{equation}
    Z = \frac{p - p_1}{\sqrt{\frac{p_1(1 - p_1)}{n}}}
\end{equation}

According to the standard normal distribution table, the critical value corresponding to an efficacy of 90\% is $Z_{0.9} = 1.28$.

In order to ensure that the efficacy is 90\%, we want to meet:

\begin{equation}
    Z_{0.9} = \frac{p_1 - p_0}{\sqrt{\frac{p_1(1 - p_1)}{n}}}
\end{equation}

The final formula to get the sample size is:

\begin{equation}
    n = \frac{p_1(1 - p_1)}{\left( \frac{p_1 - p_0}{1.28} \right)^2}
\end{equation}

The actual reject rate $p_1$ for electronic parts is known through IPC to be in the range of 0.01 to 0.10. In this paper, p1 is chosen to be in the range of 0.04 to 0.08, and a step size of 0.01 is chosen to calculate the corresponding sample size. Figure \ref{fig:Variation of sample size n for different values of $p_1$} shows the results of the calculation based on the previously derived sample size formula:

\begin{figure}[H]
  \centering
  \includegraphics[width=0.4\textwidth]{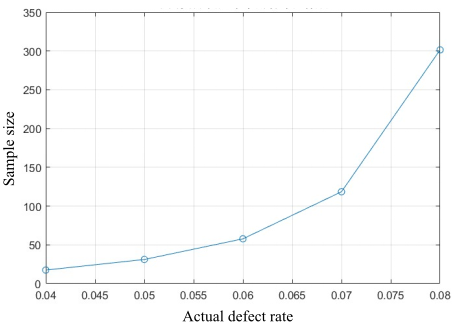}
  \caption{Variation of sample size n for different values of $p_1$}
  \label{fig:Variation of sample size n for different values of $p_1$}
\end{figure}

From the calculation results, it can be seen that when $p_1$ is close to $p_0 = 0.10$, the required sample size $n$ increases significantly, and from Figure \ref{fig:Variation of sample size n for different values of $p_1$}, it can be seen that when the actual defective rate $p_1$ is raised from 0.04 to 0.08, the required sample size $n$ is raised from 18 to 303. This is because when the actual defective rate $p_1$ is close to the defective rate claimed by the supplier, the difference between the defective rate in the samples and the nominal defective rate becomes more difficult to detect. In order to make a correct decision despite this situation, the required sample size must be large enough to improve the accuracy of the detection. When the variance is small, this variance may be masked by random fluctuations in the sample, so a larger sample size is needed to ensure that the detected defective rate accurately reflects the actual quality of the lot. This also explains why, as the actual defective rate $p_1$ gets closer to the nominal defective rate

This also explains why, as the actual defective rate $p_1$ gets closer to the nominal defective rate $p_0$, the sample size required increases significantly, even approaching infinity.

\section*{III.Dynamic Programming Modeling and Solving}

\textbf{A.Model Symbols}

To facilitate the subsequent construction of the model, we specify the notation as follows:

$r_1$,$r_2$,$r_3$: Parts 1, Parts 2, and the defective rate of the finished product.

$c_1$,$c_2$,$c_3$: cost of part 1, part 2 and assembly cost of finished product.

$t_1$,$t_2$,$t_3$: Parts 1, Parts 2, and the cost of testing the finished product.

$h_1$,$m$ : dismantling cost and replacement loss of non-conforming finished products.

$w$: market price of the finished product.

\noindent\textbf{B.Dynamic Programming}

In this paper, the main advantages of dynamic programming are:

Multi-stage decision-making model: this paper involves multiple decision-making steps in the production process (e.g., whether to detect spare parts, whether to detect finished products, whether to dismantle nonconforming products, etc.), and there is interdependence between these steps, which is in line with the multi-stage decision-making characteristic of dynamic programming.

State dependence: the decision of each stage directly affects the state of the subsequent stage, for example, the decision of detecting spare parts affects the defective rate of the finished product, and the defective rate of the finished product determines whether further detection or disassembly operations are needed. Dynamic programming finds the optimal solution between different phases by defining state transfer equations.

Global optimal solution: Dynamic programming ensures that all stages of the production process move towards the global optimal solution through step-by-step recursive derivation, avoiding the problem of local optimal solution.

To make dynamic programming simpler and clearer, we should consider how to divide the problem into a minimized number of stages while ensuring that all key decision points are covered and that costs and benefits are easy to calculate. We divide the problem into the following three phases:

\begin{enumerate}
    \item spare parts procurement and testing phase;
    \item assembly and inspection of finished products;
    \item disassembly and market flow of non-conforming finished products.
\end{enumerate}

Determine the state variables and decision variables

According to different stages, we categorize the state variables into two different forms:

$n_{11}$,$n_{12}$: denote the number of spare parts 1 and spare parts 2 in the first stage, respectively.

$n_2$,$n_3$: the number of remaining finished products in the second and third stages, respectively.

In different phases, we define different decision variables $s_k\in \left \{ 0,1 \right \} $ to denote different decision scenarios.

Specifically:

$(s_1,s_2)$: whether or not to test parts 1 and 2. (0 means no testing, 1 means testing)

$s_3$: Whether to test the assembled finished product. (0 means no testing, 1 means testing)

$s_4$: Whether to disassemble the nonconforming product. (0 means no disassembly, 1 means disassembly)

Determine the state transfer equation

Phase I to Phase II:

\setcounter{equation}{0}

\begin{equation}
    n_2=\min \left \{ n_{11}\cdot \left ( 1-r_1 \right )^{s_1} ,n_{12}\cdot \left ( 1-r_2 \right )^{s_2} \right \} 
\end{equation}

Phase II to Phase III:

\begin{equation}
    n_3=n_2\cdot \left ( 1-r_1 \right ) ^{1-s_1}\cdot \left ( 1-r_2 \right ) ^{1-s_2}\cdot \left ( 1-r_3 \right ) 
\end{equation}

List the objective functions by stage $V_{k,n}$

In this paper we define the objective function $V$ as the total profit earned by the product, which we define as:

$V_{\text{Total profit}} = V_{\text{Sales}} - V_{\text{Cost of spare parts}}$
\begin{align}
    &- V_{\text{Assembly costs}} \nonumber\\
    &- V_{\text{Cost of testing}} \nonumber\\
    &- V_{\text{dismantling costs}} \nonumber\\
    &- V_{\text{Loss on exchange}}
\end{align}

It should first satisfy the objective function divisibility, i.e., the

\begin{equation}
    V_{k,n}=v_k\left ( n_k,s_k \right ) \oplus V_{k+1,n}
\end{equation}

Our goal is to maximize profit, so we use $f_k(n_k)$ to denote the optimal objective function of the backward subprocess from $k$ to the final stage $n$ when the state of the kth stage is $n_k$, and we can obtain the following staged backward recursive equation:

\begin{equation}
    f_k\left ( n_k \right )=\max_{s_k}  \left [ v_k\left ( n_k,s_k \right )+f_{k+1}\left ( n_{k+1} \right )   \right ]
\end{equation}

\begin{figure}[H]
  \centering
  \includegraphics[width=0.4\textwidth]{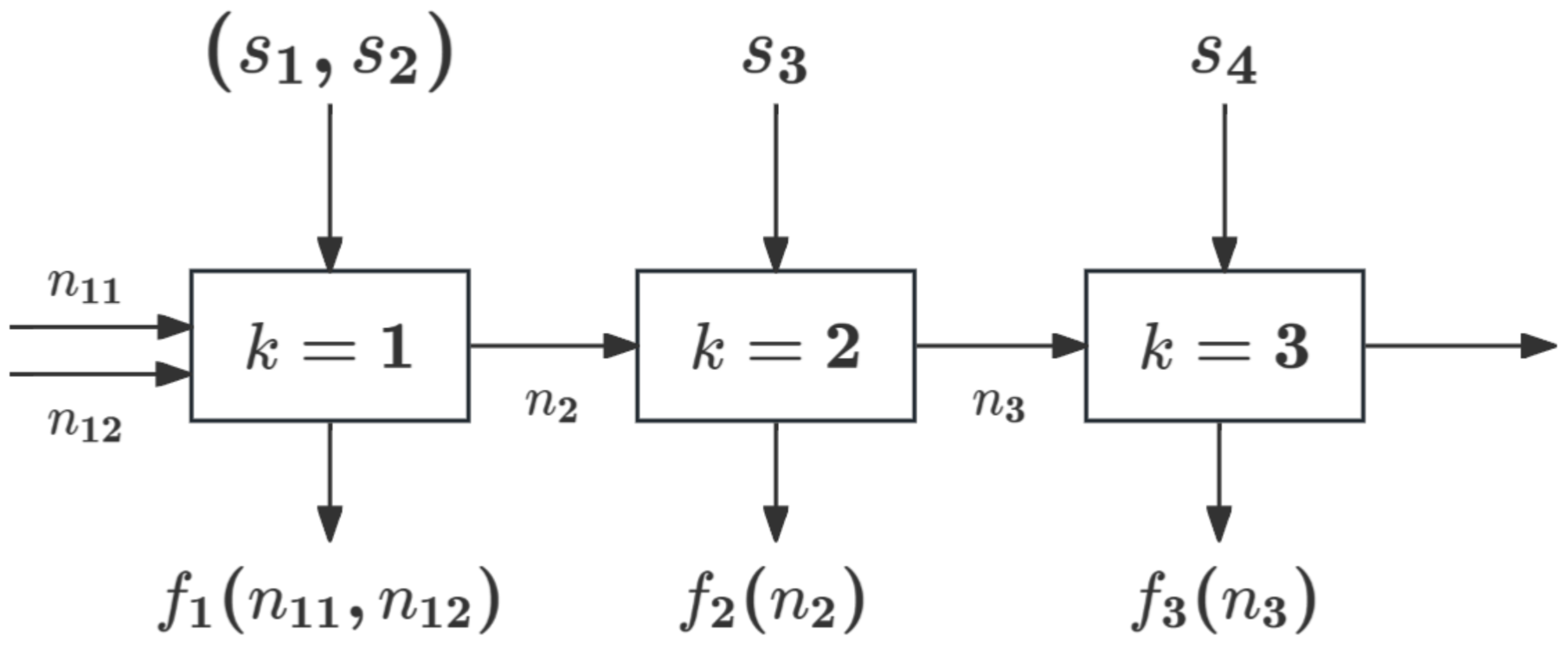}
  \caption{The inverse-order recursive process of dynamic programming}
  \label{fig:The inverse-order recursive process of dynamic programming}
\end{figure}

The recursive process is shown in Figure \ref{fig:The inverse-order recursive process of dynamic programming}.

The optimal objective function value is obtained when $f_1(n_1)$ is computed according to equation (24), and the optimal policy $s_k(k=1,2,3,4)$ is obtained by sequentially searching for the policy that is optimal when $V_{k,n}$ is reached.

When $s_4=0$, no gain occurs: the

\begin{equation}
   V_3\left ( n_3,0 \right ) =0 
\end{equation}

When $s_4=1$, i.e., the dismantling of the nonconforming finished product incurs dismantling costs, but saves the production costs of Parts 1 and 2, and we also consider the inspection costs of the dismantled parts and the assembly costs of the finished product.

\begin{equation} 
\begin{split}
    V_3\left ( n_3,1 \right ) =
    &\left ( n_2-n_3 \right ) \cdot  \left( -h_1  + \left( c_1 + c_2 \right) \right. \\
    &\left. - \left( s_1 \cdot t_1 + s_2 \cdot t_2 + s_3 \cdot \left( t_3 + c_3 \right) \right) \right) \\
    & \cdot \min \left\{ \left( 1 - r_1 \right)^{s_1}, \left( 1 - r_2 \right)^{s_2} \right\} 
\end{split}
\end{equation}

When $s_3=0$, putting the assembled finished product directly on the market will result in a partial profit and a certain amount of switching loss.
    
\begin{align}
    V_2\left ( n_2,0 \right ) =-c_3 &\cdot n_2 -r_3\cdot m\cdot n_2+  w\cdot n_2\nonumber\\&+V_{3}\left ( n_3,s_4 \right )
\end{align}

When $s_3=1$, testing the assembled product before placing it on the market is partially profitable and incurs a certain amount of testing costs.

\begin{align}
    V_2\left ( n_2,1 \right ) =-c_3 &\cdot n_2 -t_3\cdot n_2+\left ( 1-r_3\right )w\nonumber\\&\cdot n_2+V_{3}\left ( n_3,s_4 \right )
\end{align}

When $s_1=1$,$s_2=1$, consider the cost of spare parts 1 and 2 and the cost of testing.

\begin{align}
    V_1\left ( n_1,1,1 \right ) \nonumber&=- ( c_1\cdot n_{11}+c_2\cdot n_{12}+t_1\cdot n_{11}\nonumber\\&+t_2\cdot  n_{12} ) +V_2\left ( n_2,s_2 \right ) 
\end{align}

When $s_1=1$,$s_2=0$, the loss due to nonconformity of part 2 is taken into account.

\begin{align}
    V_1\left ( n_{11},n_{12},1,0 \right ) =&-( c_1\cdot n_{11}+c_2\cdot n_{12}\nonumber\\&+t_1\cdot n_{11}+r_2\cdot n_{12}\cdot  m\nonumber\\&\cdot \left ( 1-s_3  \right )  ) +V_2\left ( n_2,s_2 \right )
\end{align}

When $s_1=0$,$s_2=1$, the loss due to nonconformity of part 1 is taken into account.

\begin{align}
    V_1\left ( n_1,0,1 \right ) =&- ( c_1\cdot n_{11}+c_2\cdot n_{12}+t_2\cdot n_{12}\nonumber\\&+r_1\cdot n_{11}\cdot m\nonumber\\&\cdot \left ( 1-s_3 \right )    ) +V_2\left ( n_2,s_2 \right ) 
\end{align}

When $s_1=0$,$s_2=0$, the losses due to non-conformity of parts 1 and 2 are taken into account.

\begin{align}
    V_1\left ( n_1,0,0 \right ) =&- ( c_1\cdot n_{11}+c_2\cdot n_{12}\nonumber\\&+\left ( n_{11}\cdot r_1+n_{12}\cdot r_2 \right ) \nonumber\\&\cdot m\cdot \left ( 1-s_3 \right )   )   +V_2\left ( n_2,s_2 \right )
\end{align}

\noindent\textbf{C.model solution}

To solve the model, we used a dynamic programming algorithm to optimize the multi-stage decision-making process. First, the model calculates the state variables of each stage step by step through the state transfer equation based on the defective rate and cost data of spare parts.

In the first stage, the decision variables $s_1$ and $s_2$ for spare parts and the defective rates $r_1$ and $r_2$ are used to calculate the effective number of finished products $n_2$ to enter the next stage. in the second stage, the decision variable $s_3$ for the finished products is considered and the state transfer equation is used to calculate the final number of finished products $n_3$. 

In the third stage, the decision variable $s_4$ is used to determine whether to dismantle the substandard products, and the final revenue is calculated by combining the market selling price and dismantling cost. The finalized model for our dynamic planning is as follows:

\setcounter{equation}{0}

\begin{equation}
    \left\{\begin{matrix}{ \max f_1\left ( n_{11} ,n_{12}\right ) }
 \\n_{k+1}=T_{k}\left ( n_k,s_k \right ) 
 \\    f_k\left ( n_k \right )=\max_{s_k}  \left [ v_k\left ( n_k,s_k \right )+f_{k+1}\left ( n_{k+1} \right )   \right ]
 \\s_1,s_2,s_3,s_4 \in \left \{ 0,1 \right \} 
 \\n_{11}=n_{12}=100
\end{matrix}\right .
\end{equation}

where $f_k(n_k)$ denotes the optimal objective function of the backward subprocess from $k$ to the final stage $n$ when the state of the kth stage is $n_k$, and $s_k$,$n_k$  denote the decision variables as well as the state variables of the kth stage.

To find the optimal solution, the algorithm iterates over all possible combinations of decisions $s_1$,$s_2$,$s_3$,$s_4$ and computes the corresponding returns under each combination. The decisions at each stage are optimized from backward to forward by dynamic recursion to maximize the total return. In each step of the computation, the algorithm not only considers the inspection cost, but also combines the loss of nonconforming parts, dismantling cost, and marketing and sales revenue. In the end, the algorithm outputs the optimal inspection and dismantling strategies for different scenarios, ensuring that revenue is maximized.

Through this dynamic planning method, the multi-stage decision-making problem can be effectively dealt with, and six specific decision-making schemes for different situations are obtained, as shown in Table \ref{tab:decision_making}:

\begin{table}[h!]
\centering
\caption{Specific Decision-Making Options and Benefits in Different Situations}
\label{tab:decision_making}
\begin{tabular}{|p{0.5cm}|p{4.5cm}|p{1cm}|} 
\hline
{\footnotesize Case} & {\footnotesize Optimal Decision Scheme} & {\footnotesize Best Return} \\ \hline
1 & \( s_1 = 0, s_2 = 0, s_3 = 0, s_4 = 1 \) & 3081 \\ \hline
2 & \( s_1 = 0, s_2 = 0, s_3 = 0, s_4 = 1 \) & 3270 \\ \hline
3 & \( s_1 = 0, s_2 = 0, s_3 = 0, s_4 = 1 \) & 2361 \\ \hline
4 & \( s_1 = 0, s_2 = 0, s_3 = 1, s_4 = 1 \) & 1919 \\ \hline
5 & \( s_1 = 0, s_2 = 0, s_3 = 0, s_4 = 1 \) & 2998 \\ \hline
6 & \( s_1 = 0, s_2 = 0, s_3 = 0, s_4 = 0 \) & 2650 \\ \hline
\end{tabular}
\end{table}

\section*{IV.conclusion}

The development and application of the multi-stage decision model based on hypothesis testing and dynamic programming represent a significant advancement in the field of decision-making. Our model addresses the critical need for a systematic approach that can effectively manage the complexities and dynamics of multi-stage decision scenarios. The proposed model not only demonstrates a strong capacity for multi-stage decision-making but also enhances the interpretability and applicability of the decision-making process.

Through the rigorous control of Type I and Type II errors and the strategic application of dynamic programming, our model offers a comprehensive framework for making informed decisions that are statistically justified and optimized across multiple stages. The findings from applying this model highlight its versatility and superiority in achieving optimal decisions, showcasing its potential to be a valuable tool in various decision-making contexts.

In summary, the integration of hypothesis testing and dynamic programming within a multi-stage decision-making framework presents a novel and effective approach for tackling the challenges of complex decision scenarios. This model’s ability to provide clear, optimal decision paths through statistical rigor and strategic planning marks a significant contribution to the decision-making literature and opens up new avenues for research and application in this field.

\bibliographystyle{unsrt}  
\bibliography{main}   

\end{document}